\documentclass[english,aps,preprint,showpacs,showkeys]{revtex4}
\usepackage[T1]{fontenc}
\usepackage[latin1]{inputenc}
\usepackage{amsmath}
\usepackage{graphicx}
\usepackage{amssymb}

\makeatletter

\usepackage{babel}
\makeatother
\begin{document}

\title{Measurement of the shear strength of a charge-density wave}

\author{K. O'Neill, K. Cicak and R.E. Thorne}

\affiliation{Laboratory of Atomic and Solid State Physics, Clark Hall, Cornell
University, Ithaca, New York 14853-2501}

\begin{abstract}
We have explored the shear plasticity of charge density waves (CDWs)
in NbSe$_{3}$ samples with cross-sections having a single micro-fabricated
thickness step. Shear stresses along the step result from thickness-dependent
CDW pinning. For small thickness differences the CDW depins elastically
at the volume average depinning field. For large thickness differences
the thicker, more weakly pinned side depins first via plastic shear,
and shear plasticity contributes substantial dissipation well above
the pinning force. A simple model describes the qualitative features
of our data, and yields a value for the CDW's shear strength of approximately
$9.5\times10^{3}$~Nm$^{-2}$. This value is orders of magnitude
smaller than the CDW's longitudinal modulus but much larger than corresponding
values for flux-line lattices, and in part explains the relative coherence
of the CDW response.
\end{abstract}

\pacs{72.15.Nj,71.45.Lr,73.23.-b,74.25.Qt}

\keywords{charge density wave; shear plasticity, NbSe$_{3}$; vortex lattices,
flux-line lattices, driven periodic media}

\maketitle
The elastic and plastic properties of driven periodic media including
charge/spin density waves (CDWs/SDWs) \cite{feinberg}, flux-line
lattices in type-II superconductors \cite{blatter,marchetti,yoon}
and Wigner crystals \cite{andrei} are central to understanding their
rich dynamics and phase behavior. Impurities, dislocations and other
disorder pin the periodic medium in each case, so that a minimum or
threshold force must be applied to produce collective motion. Shear
moduli are especially important because they determine the extent
of shear plasticity. In experiments on moving flux-line lattices,
shear plasticity dominates so that the coherent oscillations at the
washboard frequency $\nu_{\lambda}$ expected in a purely elastic
system are replaced by an incoherent response with large $\frac{1}{f}$-like
noise \cite{clem}, and vanishing of the shear modulus leads to liquid-like
motion.

Shear plasticity also plays an important role in charge- and spin-density
wave motion, rounding the depinning transition, broadening the spectral
width of the coherent oscillations, and smearing out velocity steps
caused by mode locking to an applied ac drive. Experiments show that
this shear primarily results from meso/macroscopic sample inhomogeneities
\cite{maher}; rare homogeneous samples of the CDW conductor NbSe$_{3}$
show highly coherent washboard oscillations and complete harmonic
and subharmonic mode locking \cite{thorne}, suggesting that the intrinsic
CDW response is nearly elastic.

Here we show that crystals with micro-fabricated steps running along
the direction of CDW conduction show a non-monotonic variation of
their depinning force with the size of the step, due to shear along
the step. A simple model reproduces this variation, and allows us
to determine the CDW's shear strength to be $9.5\times10^{3}$~Nm$^{-2}$.
This value is more than three orders of magnitude smaller than the
shear elastic modulus \cite{dicarlo_adelman}, and two orders of magnitude
smaller than the contribution of the pinned CDW to the shear modulus
of the crystal \cite{xiang}. It is much larger than for flux-line
lattices, which explains in part the relative coherence of the CDW
response.

NbSe$_{3}$ and related quasi-one-dimensional CDW materials grow as
long thin ribbons. Shear usually results from steps in crystal thickness
associated with small-angle grain boundaries that run along the ribbon
($\mathbf{b}$) axis, which corresponds to the direction of CDW motion.
Because typical crystal thicknesses are smaller than the CDW's bulk
phase correlation length in the thickness $\mathbf{a^{\star}}$ direction
($\sim2\ \,\mu$m in undoped crystals), the depinning field varies
inversely with crystal thickness $E_{T}\propto\frac{1}{t}$ \cite{yetman,mccarten}.
As a result, thicker regions of the crystal cross-section have smaller
depinning fields than thinner regions and tend to shear away from
them, as illustrated in, e.g., Figures 3 and 4 of Li \emph{et al.}
\cite{li}. 

\begin{figure}
\begin{center}\includegraphics[%
  clip,
  scale=0.45]{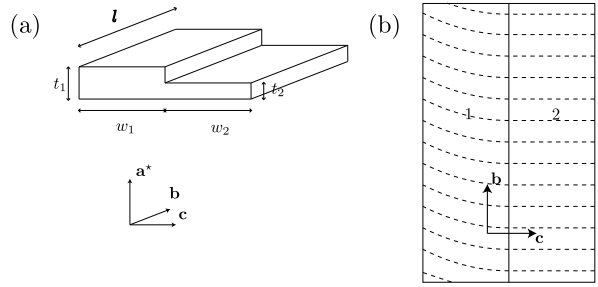}\end{center}

\caption{CDW shear in NbSe$_{3}$. {\bf (a)} A NbSe$_{3}$ single crystal
with a step running along its length. Since the pinning strength varies
inversely with thickness, the CDW in the thicker part 1 is more weakly
pinned than in the thinner part 2. {\bf (b)} When an electric field
greater than the depinning field of part 1 is applied, the CDW in
part 1 moves relative to part 2, producing shear strains.}

\label{fig:CDW-shear}
\end{figure}

Detailed information about shear can be obtained by measuring the
CDW response versus step height in samples with a single well-defined
step, shown in Figure \ref{fig:CDW-shear}. The CDWs in rectangular
regions 1 and 2 of thickness $t_{1}$ and $t_{2}$ have depinning
fields $E_{T1}$ and $E_{T2}$. They interact via a shear force $F_{shear}$
along the direction of CDW motion at the interface between them. This
force acts to retard motion of the thicker, more weakly pinned region
1 and assist motion of the thinner, more strongly pinned region 2. 

If no slip occurs at the boundary (the static friction regime), then
elastic coupling causes regions 1 and 2 depin at a common field $E_{T}$
given by \begin{equation}
E_{T,el}=\frac{E_{T1}t_{1}w_{1}+E_{T2}t_{2}w_{2}}{t_{1}w_{1}+t_{2}w_{2}},\label{eq:noslipfull}\end{equation}
where $w_{1}$ and $w_{2}$ are the widths of regions 1 and 2, respectively.
In thin crystals where $E_{T}\propto\frac{1}{t}$ and assuming $w_{1}=w_{2}$,
Eq.~\ref{eq:noslipfull} reduces to $E_{T,el}=2E_{T1}/(1+\frac{t_{2}}{t_{1}})$. 

The frictional force $F_{shear}$ has a maximum static value $F_{max}$,
beyond which plastic slip occurs along the interface. The thick region
1 shears and slides relative to the thin region 2 at a field \begin{equation}
E_{T,pl}=E_{T1}+\frac{F_{max}}{Q_{1}}\label{eq:plastic-shear}\end{equation}
 where $Q_{1}=en_{c}w_{1}t_{1}l$ is the total CDW charge in region
1 which couples to the electric field. The maximum shear force should
be proportional to the area of the interface between the two regions,
and thus to $t_{2}$. With $t_{1}$ fixed, Eq.~\ref{eq:plastic-shear}
becomes \begin{equation}
E_{T,pl}=E_{T1}+\frac{\sigma_{s}}{en_{c}w_{1}}\frac{t_{2}}{t_{1}}\label{eq:shear-strength}\end{equation}
where $\sigma_{s}$ is the plastic shear strength.

To study CDW shear, a NbSe$_{3}$ crystal with a nearly rectangular
cross-section is selected and placed on a substrate patterned with
an array of non-perturbing gold electrical probes, each $2$~$\mu$m
wide in the direction of CDW motion. At least $1$~$\mu$m of UV-5
resist is spun onto the crystal and substrate and then cured at $130^{\circ}$C,
the highest temperature to which the crystal is exposed during processing.
The resist is patterned using a Cambridge-LEICA E-beam system 10.5.
After developing, the pattern is etched into the sample using a CF$_{4}$
dry plasma etch. The etch depth and spatial consistency can be monitored
using the resistance per unit length for each probe pair along the
sample. The average thickness is then calculated from the optically
measured crystal width and NbSe$_{3}$'s room temperature resistivity
of $1.86$~$\Omega\mu$m \cite{mccarten}. More details on the fabrication
process are found in \cite{oneill}. 

\begin{figure}
\begin{center}\includegraphics[%
  clip,
  scale=0.45]{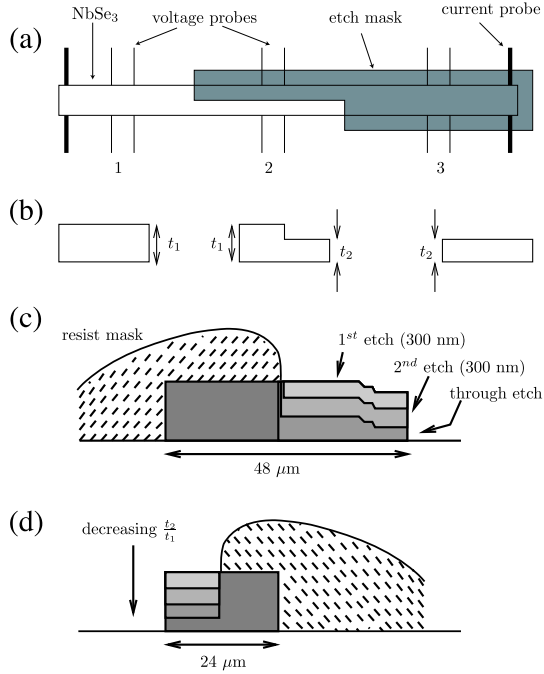}\end{center}

\caption{{\bf (a)} Sample A.  Etching thinned the sample region not covered
by the mask, and the resulting crystal cross sections in sample segments
1, 2, and 3 are shown in {\bf (b)}. The sample width was $62$~$\mu$m,
its unetched thickness was $t_{1}=0.52$~$\mu$m, and the etched
region thickness was $t_{2}=0.26$~$\mu$m. {\bf (c)} and {\bf (d)}
Sample B. The step is etched through a resist mask along the entire
sample length. The unetched sample was $48$~$\mu$m wide and $0.89$~$\mu$m
thick. After the first etch sequence in (c), the entire crystal was
reduced to half its original width, and preexisting growth steps removed.
 A second etch sequence (d) on the rectangular sample produced in
(c) yielded the most accurate measurements of depinning field versus
$\frac{t_{2}}{t_{1}}$.}

\label{fig:step-sample}
\end{figure}

We explored two different sample designs. The first, shown in Figure
\ref{fig:step-sample}(a) and (b), has three distinct cross-sections:
an as-grown rectangular segment at one end with thickness $t_{1}$,
an etched rectangular segment at the other end with thickness $t_{2}$,
and a middle segment with a step running along $\mathbf{b}$ separating
unetched and etched regions of thickness $t_{1}$ and $t_{2}$, respectively.

The second sample design, shown in Figure \ref{fig:step-sample}(c)
and (d), is etched to produce a thickness step along its entire length.
This eliminates current density changes and contributions to $E_{T}$
from longitudinal phase slip \cite{gill_maher} present in the first
design, and thus yields more precise measurements of shear effects
on $E_{T}$. The NbSe$_{3}$ crystal used had two small steps. A portion
of the crystal width containing the steps was thinned by repeated
etching through the same resist mask, as shown in Figure \ref{fig:step-sample}(c),
and transport measurements were performed after each etch. This yielded
$E_{T}$ as a function of the thickness ratio $\frac{t_{2}}{t_{1}}$.
This process was then repeated using the remaining, initially step-free
part of the crystal, as shown in Figure \ref{fig:step-sample}(d).
Data for two samples, A and B, fabricated using the first and second
process, respectively, are described here. Measurements were performed
at $T=120$~$K$ where longitudinal phase slip contributions to $E_{T}$
are smallest. For both samples the thickness was less than the bulk
CDW phase-phase correlation length in the $\mathbf{a^{\star}}$ direction,
so that depinning fields $E_{T}\propto1/t$ are expected.

\begin{figure}
\begin{center}\includegraphics[%
  clip,
  scale=0.3]{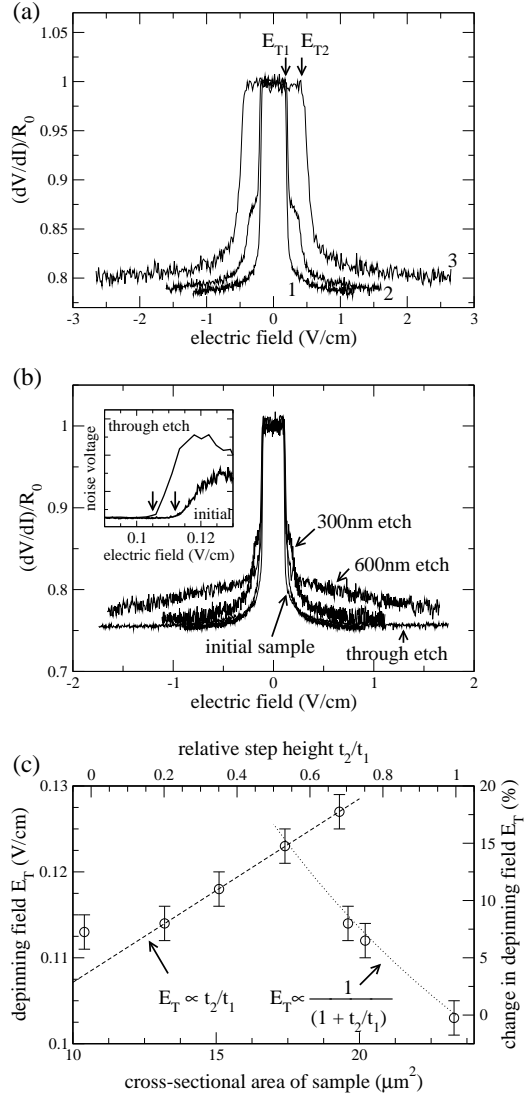}\end{center}

\caption{{\bf (a)} Normalized $\frac{dV}{dI}$ versus electric field at $T=120$~K
for Sample A. Curves 1, 2 and 3 were measured between voltage probes
on segments 1, 2 and 3, in Figure \ref{fig:step-sample}(a), respectively.
$\frac{dV}{dI}$ drops abruptly from the low-field single particle
resistance at the collective depinning field $E_{T}$. {\bf (b)}
Normalized $\frac{dV}{dI}$ and versus electric field at $T=120$~$K$
for Sample B, acquired during the first etch sequence in Figure \ref{fig:step-sample}(c).
The inset shows the noise amplitude, measured in a $4$~Hz bandwidth
around $20$~Hz, that abruptly increases at $E_{T}$. {\bf (c)}
Depinning field $E_{T}$ (from noise measurements) versus thickness
ratio $t_{2}/t_{1}$ and cross-sectional area, obtained for Sample
B during the second etch sequence of Figure \ref{fig:step-sample}(d).}

\label{fig:all-data}
\end{figure}

Figures \ref{fig:all-data}(a) and \ref{fig:all-data}(b) show the
measured four-probe differential resistance $\frac{dV}{dI}$ versus
electric field on Samples A and B, respectively. For Sample A, the
unetched region 1 and uniformly etched region 3 have depinning fields
$E_{T}$ in the ratio $1:2.4$. This roughly matches the thickness
ratio $1:2.0$ determined from room temperature resistance measurements.
Consequently, despite producing a somewhat rougher surface, etching
yields roughly the same thickness-dependent $E_{T}$ as is observed
in unetched crystals of different thicknesses \cite{mccarten}. Curve
2 in Figure \ref{fig:all-data}(a) for the stepped segment 2 shows
two separate depinnings for the etched and unetched portions of the
crystal cross-section. The smaller depinning field of the unetched
portion is increased only slightly from that of the unetched and step-free
segment 1, suggesting that it shears from the etched portion. The
larger depinning field of the thin, etched portion is lower than that
of the unstepped etched segment 3, indicating that shear forces exerted
by the depinned CDW in the thick portion assist depinning in the thin
portion. 

For Sample B, the normalized resistivities versus field in Figure
\ref{fig:all-data}(b) for the initial, unetched sample and for the
through-etched sample are nearly identical and show a large, abrupt
drop at $E_{T}$, indicating excellent sample quality and that side-wall
roughness induced by etching is unimportant. In contrast, for partial
etching the differential resistance well above $E_{T}$ is significantly
larger than for the unetched or through-etched sample, and grows with
etch depth. The differential conductance associated with CDW motion
at $E=16E_{T}$ is 9\% smaller for a $600$~nm etch depth ($t_{2}/t_{1}=0.33$)
than for the through-etched crystal. This indicates that shear friction
- in the form of shear-induced CDW phase vortex/dislocation ``turbulence''
- along the step strongly affects CDW dynamics.

To accurately determine how the depinning field of Sample B varies
with etched thickness ($t_{2}/t_{1}$), both the low-frequency noise
amplitude and the differential resistance were measured versus electric
field, as shown in Figure \ref{fig:all-data}. The noise amplitude
increases by orders of magnitude when the CDW or a portion of it depins
\cite{fleming}, and provides the most sensitive probe of $E_{T}$
. The cleanest data were obtained for the second etch sequence of
Figure \ref{fig:step-sample}(d), starting with the rectangular cross-section
produced in \ref{fig:step-sample}(c). To simplify interpretation,
the widths of the etched and unetched portions were chosen to be nearly
identical and equal to $12$~$\mu$m.

Figure \ref{fig:all-data}(c) shows the depinning field $E_{T}$ versus
both sample cross-sectional area and thickness ratio $\frac{t_{2}}{t_{1}}$.
Although there is some scatter, the data clearly show two distinct
regimes. For $\frac{t_{2}}{t_{1}}$ close to $1$ (small steps), $E_{T}$
increases with decreasing $\frac{t_{2}}{t_{1}}$. For $\frac{t_{2}}{t_{1}}$
close to $0$ (large steps), $E_{T}$ increases with increasing $\frac{t_{2}}{t_{1}}$.
This general behavior is consistent with the simple model of Equations~\ref{eq:noslipfull}-\ref{eq:shear-strength}.
For large etch thicknesses $\frac{t_{2}}{t_{1}}\lesssim1$, the shear
stress at the boundary between etched and unetched regions is small.
$\frac{dV}{dI}$ drops sharply and smoothly, indicating that the entire
sample volume depins at a unique $E_{T}$, determined by the volume
average of $E_{T1}$ and $E_{T2}$ given by Equation \ref{eq:noslipfull}.
As the etched thickness and $\frac{t_{2}}{t_{1}}$ decrease, the shear
stress $\sigma_{s}$ eventually is sufficient to cause shear slip
along the boundary. In this case, $\frac{dV}{dI}$ shows two successive
``drops'' corresponding to depinning of the unetched (thick) and etched
(thin) regions, respectively. The thick region's depinning field $E_{T}$
is the sum of the pinning force $\propto E_{T1}$ in that region and
the maximum retarding shear force exerted across the area of contact
between the CDWs. This is determined by the maximum shear stress $\sigma_{s,max}$
(force/area) times the contact area $\propto t_{2}$ and is given
by Equation \ref{eq:plastic-shear}.

The dotted and dashed lines in Figure \ref{fig:all-data}(c) are fits
to Eqns. \ref{eq:noslipfull} and \ref{eq:plastic-shear} in the elastic
and plastic regimes, respectively. The overall qualitative agreement
is remarkable given the simplicity of our model. The elastic regime
fit has no adjustable parameters, since $E_{T}(t)$ for our undoped
NbSe$_{3}$ 
crystals has been determined in independent measurements \cite{mccarten},
and is in good quantitative agreement as well. The CDW's elastic shear
stress along the boundary grows as $t_{2}/t_{1}$ becomes smaller.
From the location of the cross-over between the elastic and plastic
regimes we can estimate the CDW's maximum elastic shear stress from
Equation \ref{eq:shear-strength} as $\sigma_{s}=\frac{en_{c}w_{1}}{2}\frac{t_{1}}{t_{2}}[E_{T}(\frac{t_{2}}{t_{1}})-E_{T1}]$.
Using the known cross-section dimensions $w_{1}$, $t_{1}$ and $t_{2}$,
a condensate density $n_{c}=1.9\times10^{21}$~cm$^{-3}$, and the
value of $E_{T}$ from the elastic fit at $t_{2}/t_{1}=0.66$ yields
$\sigma_{max,el}=6.6\times10^{3}$~Nm$^{-2}$. The plastic fit (dashed
line) in Figure \ref{fig:all-data}(c) is given by $E_{T}(\frac{t_{2}}{t_{1}})-E_{T}(0)\approx0.026\frac{t_{2}}{t_{1}}$
(in Vcm$^{-1}$). From this fit we obtain the plastic shear strength
$\sigma_{s}$ at the interface between the thick and thin portions
of the sample of $\sigma_{s}\approx9.5\times10^{3}$~Nm$^{-2}$,
in rough agreement with the maximum elastic shear strength. These
estimates do not account for the effects of non uniform stresses along
the interface due to the L-shaped cross-section. 

A rough lower bound for the CDW's shear modulus $C$ may be estimated
from the ratio of the maximum elastic shear stress to the CDW shear
strain when adjacent chains are $\frac{\pi}{2}$ out of phase. This
gives the relation $\sigma_{s}\leq\frac{1}{4}\frac{\lambda_{c}}{d}C$,
where $\lambda_{c}$ is the CDW wavelength and $d$ is the distance
between adjacent chains \cite{smaalen}. This gives $C\geq2.1\times10^{4}$~Nm$^{-2}$,
more than $3$ orders of magnitude smaller than NbSe$_{3}$'s measured
longitudinal modulus $\sim5\times10^{8}$~Nm$^{-2}$ \cite{dicarlo_adelman}
by the measured ratio $\sim10$ of the longitudinal and transverse
correlation lengths. As in ordinary crystals and flux-line lattices,
the measured stress is likely reduced from its theoretical value by
crystal and CDW defects.

The elastic moduli of flux-line lattices depend on magnetic field.
The shear modulus $c_{66}$ increases roughly linearly at small fields,
reaching a maximum at intermediate fields before dropping to zero
at high fields. In Nb$_{3}$Ge, the maximum value is $\sim10$Nm$^{-2}$
at $\frac{B}{B_{c2}}\sim0.3$ \cite{pruymboom}. Measurements on vortices
in nm-scale, weakly pinned Nb$_{3}$Ge channels in a strongly pinned
background \cite{pruymboom} yield a maximum plastic shear strength
at $\frac{T}{T_{c}}=0.6$ and $\frac{B}{B_{c2}}\sim0.4$ of $\sim0.4$~Nm$^{-2}$.
These shear modulus and shear strength values are roughly six and
four orders of magnitude smaller, respectively, than the corresponding
CDW values, consistent with the much more nearly elastic collective
response of CDWs. The character of the bulk collective response depends
on the ratio of the elastic strength to the bulk pinning strength,
which is reflected in the ratio of the correlation length to the lattice
periodicity. In CDW systems this ratio is typically $10^{3}$ to $10^{4}$,
two to three orders of magnitude larger than is typical in flux-line
lattices.

In conclusion, we have observed a crossover from elastic to plastic
behavior in the depinning of charge-density waves in crystals with
artificially produced cross-section steps. From this behavior we have
determined the shear strength of the $T_{P1}$ CDW in NbSe$_{3}$.
This measurement provides a basis for understanding shear plasticity
in CDW systems, and its relation to plasticity observed in other driven
disordered media. 

\begin{acknowledgments}
We thank Rut Besseling for fruitful discussions. This work was supported
by the National Science Foundation (NSF) (DMR 0101574 and INT 9812326).
K. O'N. was supported by a U. S. Department of Education Fellowship.
Nanofabrication work was performed at the Cornell Nano-Scale Science
\& Technology Facility, supported by the NSF(ECS-9731293), its users,
and its industrial affiliates.
\end{acknowledgments}

\end{document}